\documentclass[prb,twocolumn,showpacs]{revtex4}   
\usepackage{graphicx}
\begin{document}

\title{
Low-energy excitations of the Hubbard model on the Kagom\'e lattice}

\author{
Yoshiki Imai
}

\email[E-mail address: ]{imai@issp.u-tokyo.ac.jp}
\author{Norio Kawakami
}

\affiliation{
Department of Applied Physics, 
Osaka University, Suita, Osaka 565-0871,Japan}

\author{
Hirokazu Tsunetsugu
}

\affiliation{
Yukawa Institute for Theoretical Physics, Kyoto University, Kyoto 606-8502, Japan}

\date{\today}

\begin{abstract}
The Hubbard model on the Kagom\'e lattice is investigated 
in a metallic phase at half-filling. 
By introducing anisotropic electron hopping on the lattice, 
we control geometrical frustration and 
clarify how the lattice geometry affects physical properties. 
By means of the fluctuation exchange (FLEX) approximation, 
we calculate the spin and charge susceptibilities, 
the one-particle spectral function, 
the quasi-particle renormalization factor, and the Fermi velocity. 
It is found that geometrical frustration 
of the Kagom\'e lattice suppresses the 
instability to various ordered states through the strong reduction of
the wavevector dependence of susceptibilities,
thereby stabilizing the 
formation of quasi-particles 
due to the almost isotropic spin fluctuations in the Brillouin zone.
These characteristic properties are discussed in 
connection with the effects of geometrical frustration 
in the strong coupling regime. 
\end{abstract}

\pacs{71.10.-w, 71.27.+a}%

\maketitle
\section{Introduction}
Geometrically frustrated metallic systems have attracted much 
interest since the discovery of the heavy fermion behavior 
in the transition metal oxide, ${\rm LiV_{2}O_{4}}$\cite{Kon97,Ura00}. 
This compound has the spinel structure and 
the band calculations show that the conduction bands are
essentially composed of vanadium 3$d$ orbitals, which
are well separate from the oxygen 2$p$ bands. 
Therefore when low-energy properties are concerned, it is sufficient
to consider only the vanadium 3$d$ orbitals \cite{Ani99,Eye99,Mat99,Sin99}.
The vanadium sublattice in the spinel structure constitutes
a network of corner-sharing tetrahedra, i.e.,
the pyrochlore lattice, and this is a typical example 
of geometrically frustrated lattices in three dimensions (3D).
Various works have revealed unusual low-energy 
properties of spin systems on the pyrochlore lattice, including
the presence of thermodynamic degeneracy of the ground states 
\cite{Lie86,Can98,Iso98,Yam00,Kog01,Tsu01A,Tsu01B}.
Another metallic pyrochlore compound, ${\rm Y(Sc)Mn_{2}}$, 
also exhibits the heavy fermion behavior 
where unusual feature has been observed in the dynamical susceptibility 
\cite{Shi93,Bal96}. 
In both compounds, no long-range order has been observed 
and the specific heat coefficient is strongly enhanced at low temperatures, 
similar to the lanthanide or actinide heavy-fermion systems. 

The heavy fermion behavior in the lanthanide or actinide systems 
is, as well known, attributed to the Kondo effect and 
the presence of localized $f$-orbital is essential to this 
mechanism \cite{Kur00}. 
However, in these transition-metal heavy fermion 
systems, 3$d$ electrons are much more mobile than $f$-electrons and 
the presence of local magnetic moment has not been detected 
so far. It is highly nontrivial whether the enormous mass 
enhancement is also attributed to the Kondo effect, 
and there may exist another mechanism leading to the heavy 
fermion-like behavior. 
This indeed gives rise to a number 
of theoretical proposals on the mechanism of the formation 
of heavy quasi-particles. 
Among them, it has been claimed
that geometrical frustration plays an important role even in the metallic
phase when the electron interactions are sufficiently large 
\cite{Iso00,Fuj01,Fuj02,Lac01,Tsu02,Bur02,
Yam02,Yod99,Cap01,Kas01,Mor02,Ima02}. 

The two-dimensional (2D) Kagom\'e lattice is 
another geometrically frustrated system, 
which may be regarded as a 2D analog of the pyrochlore lattice. 
Antiferromagnetic spin systems on this lattice have been intensively studied 
and many unusual properties have been found.
For example, in the $S=1/2$ Heisenberg antiferromagnet, there exists 
a finite energy gap between the singlet ground state 
and triplet excitations, 
and a thermodynamic number of singlet excitations exist 
within the singlet-triplet gap 
due to strong frustration\cite{Els89,Els95,Cha92,Mil98,Sin00}. 
This level scheme of excitations is in common with that for 
the spin systems on the 3D pyrochlore lattice, and it may 
be quite generic in geometrically frustrated systems. 
If electrons become itinerant, electron motion 
will be coupled to both spin-triplet and spin-singlet 
excitations. Therefore, quasi-particles in this case 
may be renormalized in a different manner 
than other more conventional cases without frustration. 
To study this issue, we will investigate the 
correlation effects in the 2D Kagom\'e lattice. 
The Kagom\'e lattice is simpler than the pyrochlore lattice 
due to the low dimensionality, but otherwise its magnetic properties 
in the insulating phase are in common with those in the pyrochlore lattice. 
Therefore, the study on a metallic Kagom\'e system is a good starting point 
to investigate how geometrical frustration affects physical properties 
in a metallic phase, and we expect that many of the results will hold 
in the other frustrated systems including the pyrochlore system. 

In this paper, we will investigate the effects of geometrical frustration of
the Kagom\'e lattice on physical properties. 
In particular, we focus on the possibility of magnetic instability 
and its relationship to the nature of quasi-particles in a metallic phase. 
We employ the fluctuation exchange (FLEX) approximation 
\cite{Bic89,Dah95,Wer96,Lan95,Tak97,Ari00} 
for electron correlations and calculate the spin and charge 
susceptibilities, and the one-particle spectral function. 
We will show that geometrical frustration of the Kagom\'e lattice 
indeed suppresses the 
instability to various ordered states, and stabilizes 
the formation of quasi-particles.

This paper is organized as follows. In the next section, 
the model and the method are described briefly, and we show 
the obtained results 
in Sec. III. Brief summary is given in Sec. IV. 

\section{Model and Method}

\begin{figure}[b]
\includegraphics[width=8cm]{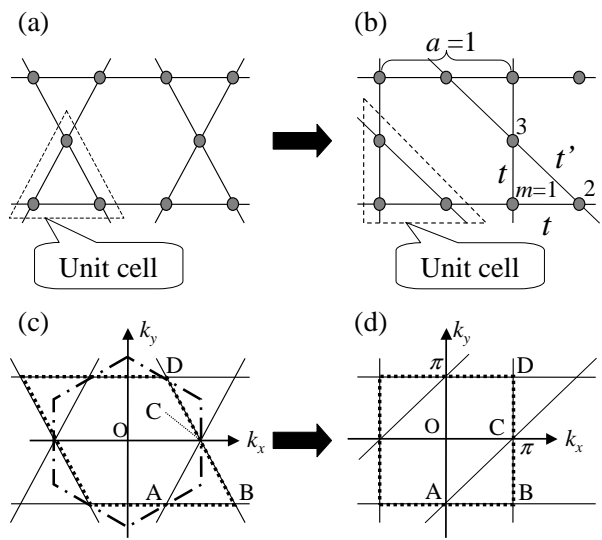}
\caption{(a) Original Kagom\'e lattice, and
(b) decorated square lattice 
which is topologically equivalent 
to (a). Lattice points labeled by $m=1$, $2$, 
and $3$ represent relative positions 
in each unit cell. $t$ and $t'$ are hopping integrals. 
(c) and (d) BZ of the original Kagom\'e 
and decorated square lattices, respectively. 
A-D represent the corresponding wavevectors in both lattices. 
Hexagon shown by dash-dotted line in (c) is 
the 1st BZ of the Kagom\'e lattice, 
which is equivalent to the diamond shown by dotted line. 
Square shown by dotted line in (d) represents 
the 1st BZ corresponding to the diamond in (c). 
}
\label{fig:lattice}
\end{figure}

The original Kagom\'e lattice is schematically 
shown in Fig. \ref{fig:lattice} (a),
which is given by a corner-sharing 2D network of triangles.
This is topologically equivalent to the decorated square
lattice with specific diagonal bonds, as 
shown in Fig. \ref{fig:lattice} (b), 
where the unit cell contains three sites. 
In the following discussions we will deal with the latter lattice, 
which makes the analysis simpler, 
since its Brillouin zone (BZ) becomes from hexagon 
(Fig. \ref{fig:lattice} (c)) 
to square, $-\pi/a <k_{x},k_{y} <\pi/a$ (Fig. \ref{fig:lattice} (d)), 
where $a$ is the lattice constant. 
In order to control geometrical frustration on the Kagom\'e lattice, 
we here introduce anisotropic hopping, 
$t$ and $t'$, between nearest neighbor sites 
as shown in Fig. \ref{fig:lattice} (b). 
When $t'/t=1.0$, the system is equivalent to the original Kagom\'e lattice. 
The advantage of introducing 
this anisotropy is that it properly interpolates the non-frustrated 
lattice (small $t'/t$) to the fully-frustrated lattice ($t'/t=1$) 
in the strong coupling limit $U/t \rightarrow \infty $ at half filling, 
where the system is reduced to the antiferromagnetic Heisenberg model. 
We expect that the effects due to the frustrated lattice geometry 
should show up even at weak or intermediate values of $U/t$. 
Hereafter we call the decorated square lattice 
with $t'/t=1.0$ ($t'/t \ne 1.0$)
{\it isotropic Kagom\'e} ({\it anisotropic Kagom\'e}) lattice 
and the lattice constant $a$ is taken to be unity, $a=1$. 

Then we consider the Hubbard model 
on the isotropic/anisotropic Kagom\'e lattice 
of Fig. \ref{fig:lattice} (b).
 The Hamiltonian is given by 
\begin{eqnarray}
H&=&\sum_{{\bf k},m,m',\sigma}\epsilon^{mm'}_{{\bf k}}
c^{\dag}_{{\bf k}m\sigma}c_{{\bf k}m'\sigma}
-\mu\sum_{{\bf k},m,\sigma}c^{\dag}_{{\bf k}m\sigma}
c_{{\bf k}m\sigma}\nonumber \\
&+&U\sum_{i,m}c^{\dag}_{im\uparrow}c_{im\uparrow}
c^{\dag}_{im\downarrow}c_{im\downarrow}, 
\label{Hamiltonian}
\end{eqnarray}
where $c_{{\bf k}m\sigma}$ ($c^{\dag}_{{\bf k}m\sigma}$) 
represents the annihilation (creation) operator of an electron at 
wave vector ${\rm \bf k}$ with site index $m$ and 
spin $\sigma$. Note that the lattice points are labeled by the 
position of each unit cell $i$ together with the relative 
position $m(=1,2,3)$ in the cell, there exist three bands. 
Here $\mu$ is the chemical potential 
and $U$ is the on-site Coulomb repulsion. 
The kinetic energy $\epsilon^{mm'}_{{\bf k}}$ 
is given in the matrix form, 
\begin{eqnarray}
\hat{\epsilon}_{{\bf k}}=
\left(
\begin{array}{ccc}
0&-2t\cos\big(\frac{k_{x}}{2}\big)&-2t\cos\big(\frac{k_{y}}{2}\big)\\
-2t\cos\big(\frac{k_{x}}{2}\big)&0
&-2t'\cos\big(\frac{k_{x}-k_{y}}{2}\big)\\
-2t\cos\big(\frac{k_{y}}{2}\big)
&-2t'\cos\big(\frac{k_{x}-k_{y}}{2}\big)&0\\
\end{array}
\right). \nonumber \\
\label{eqn:S4-dispersion}
\end{eqnarray}

In the following discussions, 
one hopping integral is fixed and taken as units of energy, $t=1$. 
In the present study, we focus on the half-filling case, 
in which geometrical frustration in the strong coupling regime is 
most prominent
and discuss how the system changes 
its characteristic properties 
from the weak coupling to the strong-coupling regime. 

In order to study electron correlations in the 
model, we employ the FLEX approximation, which 
is a self-consistent perturbation method 
with respect to the Coulomb interaction $U$. 
The FLEX approximation is a conserving 
approximation based on the idea of Baym and Kadanoff \cite{Bay61,Bay62}, 
and has been successfully used to describe 
electron correlations in the high-$T_{c}$ cuprates 
and other correlated electron systems
\cite{Bic89,Dah95,Wer96,Lan95,Tak97,Ari00}. 
In our three-band system, it is convenient to 
represent the Green function $\hat{G}$ and the 
effective interaction $\hat{V}$ in the $3 \times 3$ matrix form 
corresponding to $3$ sites in the unit cell. 
As far as there is no spin order, the Green function and other quantities 
are diagonal in spin space, and we drop the spin index. 
The self energy within the FLEX approximation is written as 
\begin{eqnarray}
\hat{\Sigma}(k)=\frac{T}{N'}\sum_{q}{}^{t}\hat{V}(-q)\hat{G}(k+q), 
\label{eqn:Sigma}
\end{eqnarray}
where $T$ is the temperature, $k \equiv ({\bf k},i\omega_n)$ and 
$q \equiv ({\bf q},i\nu_{l})$, with $\omega_n=(2n-1)\pi T$ 
and $\nu_l=2l\pi T$, and $N'=N/3$ is the number of total unit cells, with
 $N$ being the number of total sites. The effective interaction is given by 
\begin{eqnarray}
\hat{V}(q)&=&U\hat{I}+U^{2}\hat{\chi}(q)+\frac{3}{2}U^{2}\hat{\chi}(q)
\Big[\Big(\hat{I}-U\hat{\chi}(q)\Big)^{-1}-\hat{I}\Big]\nonumber \\
&&+\frac{1}{2}U^{2}\hat{\chi}(q)\Big[\Big(\hat{I}+U\hat{\chi}(q)
\Big)^{-1}-\hat{I}\Big]\\
\hat{\chi}(q)&=&\frac{-T}{N'}\sum_{k}\hat{G}(k+q)\hat{G}(k), 
\end{eqnarray}
where $\hat{I}$ is the unit matrix. 
The Dyson equation for the renormalized Green function reads
\begin{eqnarray}
\hat{G}(k)^{-1}&=&\hat{g}(k)^{-1}-\hat{\Sigma}(k),
\label{eqn:Dyson}
\end{eqnarray}
where $\hat{g}(k)$ is the bare Green function defined as 
\begin{eqnarray}
\hat{g}(k)&=&\Big[(i\omega_n+\mu)\hat{I}-
\hat{\epsilon}_{\bf k}\Big]^{-1}.
\end{eqnarray}
By numerically iterating the procedures of 
Eqs. (\ref{eqn:Sigma})-(\ref{eqn:Dyson}), 
we obtain the renormalized Green function. 

In order to investigate the spin and charge 
response of the system, we introduce the 
generalized susceptibility $\hat{\chi'}(q)$ 
of which element is defined by 
\begin{eqnarray}
&&\chi'_{m_{1}m_{2},m_{3}m_{4}}(q)\nonumber \\
&&=\int_{0}^{\beta}{\rm d}\tau \,{\rm e}^{i\nu_{l} \tau}\,
\langle \rho_{m_{1}m_{2}}({\bf q},\tau)\rho^{\dag}_{m_{3}m_{4}}
({\bf q},0)\rangle,
\end{eqnarray}
where $\rho_{mm'}({\bf q},\tau)\equiv
(1/\sqrt{N'})\sum_{i}{\rm e}^{i{\bf q}\cdot{\bf R}_{i}}
c^{\dag}_{im}(\tau)c_{im'}(\tau)$ is 
the generalized polarization at imaginary time $\tau$. 
${\bf R}_{i}$ represents the position
of the unit cell $i$. 
It should be noted that generalized susceptibility $\hat{\chi'}$ 
is in the $9 \times 9$ matrix form 
and is diagonal for the spin sector. 
Within the FLEX approximation, the generalized 
susceptibility is obtained as
\begin{eqnarray}
&&\chi'_{m_{1}m_{2},m_{3}m_{4}}(q)\nonumber \\
&=&\frac{-T}{N'}\sum_{k}
G_{m_{2}m_{4}}(k+q)G_{m_{3}m_{1}}(k)\nonumber \\
&\times&{\rm e}^{{-i{\bf k}\cdot({\bf r}_{m_{1}}-{\bf r}_{m_{3}})}}
{\rm e}^{{i({\bf k}+{\bf q})\cdot({\bf r}_{m_{2}}-{\bf r}_{m_{4}})}}, 
\label{eqn:chi'}
\end{eqnarray}
where ${\bf r}_{m}(m=1$-$3)$ is a relative lattice position in each 
unit cell (Fig. \ref{fig:lattice}(b)). 
The spin and charge susceptibilities, 
$\hat{\chi}^{(s)}$ and $\hat{\chi}^{(c)}$, 
are the response of spin-triplet and spin-singlet polarizations, respectively, 
and their matrix element is given by 
\begin{eqnarray}
&&\chi^{\rm (s,c)}_{m_{1}m_{2},m_{3}m_{4}}(q) \nonumber \\
&=&\Big[\hat{\chi'}(q)
\Big(\hat{I}\mp\hat{U}\hat{\chi'}(q)\Big)^{-1}\Big]
\Big|_{m_{1}m_{2},m_{3}m_{4}}, 
\end{eqnarray}
where the $\mp$ sign should read $-$ for the spin and $+$ for the charge 
susceptibilities, respectively. The element of $\hat{U}$ is
\begin{eqnarray}
U_{m_{1}m_{2},m_{3}m_{4}}=U\delta_{m_{1},m_{2}}\delta_{m_{3},m_{4}}\delta_{m_{1},m_{3}}. 
\end{eqnarray}

\section{Results}
We numerically iterate the above-mentioned procedure 
of Eqs. (\ref{eqn:Sigma})-(\ref{eqn:Dyson}) 
until the calculated $\hat{G}(k)$'s converge within desired accuracy. 
The summations are efficiently carried out by using the fast 
Fourier transform (FFT) with $N'=64^{2}$ points 
in the ${\bf k}$-summation 
and 2048 Matsubara frequencies in the $\omega_{n}$-summation . 

\subsection{Free Electron Properties}

We start with the non-interacting case, $U/t=0$. By using an orthogonal 
transformation, the kinetic term of the Hamiltonian is diagonalized 
at each ${\bf k}$. The result for the isotropic case $t'/t=1.0$ is 
given by 
\begin{eqnarray}
E_{{\bf k}1,2}&=&
-t\bigg[1\pm
\sqrt{1+8\cos\Big(\frac{k_{x}}{2}\Big)\cos\Big(\frac{k_{y}}{2}\Big)
\cos\Big(\frac{k_{x}-k_{y}}{2}\Big)}\bigg] \nonumber \\
E_{{\bf k}3}&=&2t,
\end{eqnarray}
where $+$ and $-$ sign correspond to the lowest band, $E_{{\bf k}1}$,
and the middle band, $E_{{\bf k}2}$, respectively.
While the largest eigenvalue forms a flat band over the whole BZ, 
the lower two dispersive bands are symmetric with respect to 
$\omega=-1.0$, and touch each other at ${\bf k}=\pm(2\pi/3,-2\pi/3)$ 
with linear dispersions. 
\begin{figure}[b]
\includegraphics[width=8cm]{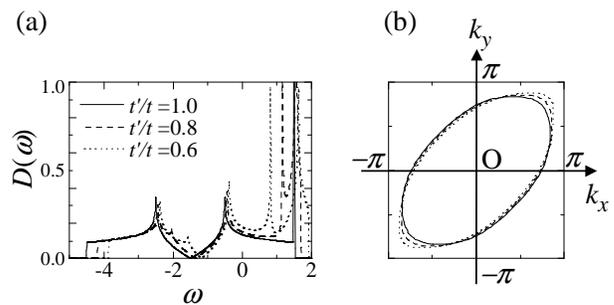}
\caption{(a) DOS and (b) Fermi surfaces 
of the non-interacting system ($U/t=0$) at half filling 
for various hopping amplitude; 
$t'/t=1.0$ (solid line), $0.8$ (dashed line) 
and $0.6$ (dotted line). 
}
\label{fig:DOS&FS}
\end{figure}
Let us define the density of states (DOS) with including
the chemical potential, 
as $D(\omega)=(1/N)\sum_{{\bf k},\alpha}1/(\omega+\mu-E_{{\bf k}\alpha})$ 
and they are shown in Fig. \ref{fig:DOS&FS} (a) for several $t'$. 
For the isotropic case $t'/t=1.0$, a $\delta$-function peak
appears at $\omega \sim 1.5$ $(\omega+\mu=2)$ 
due to the flat band $E_{{\bf k}3}$.
In the anisotropic case, the dispersion of the three bands
is modified, and the main change is that the highest band
is now dispersive, leading to the broadening of the
$\delta$-function part of the DOS. However, the qualitative
change in the DOS around the Fermi level, $\omega=0$, is quite small.
Shown in Fig. \ref{fig:DOS&FS} (b) is 
the Fermi surface for the corresponding cases. 
When $t'/t=1.0$, the shape of the Fermi surface is most isotropic. 
As $t'$ decreases, the Fermi surface is elongated 
along the $k_{y}=k_{x}$ direction, and becomes more anisotropic. 

\begin{figure}[tb]
\includegraphics[width=8cm]{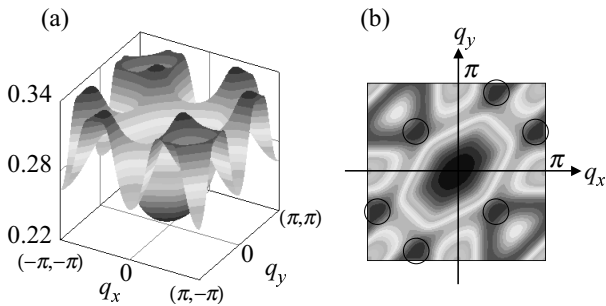}
\caption{(a) Largest eigenvalues of the static
 susceptibility $\hat{\chi'}({\bf q},0)$ at
half-filling for $U/t=0$. (b) Contour plot of (a). 
The circles indicate the 
location of the peak values. The temperature is $T/t=0.1$. 
}
\label{fig:EVchiU0}
\end{figure}

\begin{figure}[tb]
\includegraphics[width=8cm]{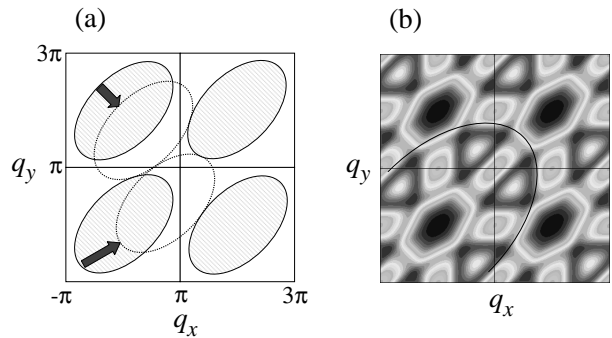}
\caption{(a) 
Nesting of Fermi surface. Each shaded ellipse depicts a Fermi surface 
in the BZ shown by a square box. Arrows show the nesting vectors. 
(b) Contour plot of the largest eigenvalues of 
$\hat{\chi'}({\bf q},0)$. The curve is a guide to eyes. 
}
\label{fig:Nesting}
\end{figure}

Once the free electron Hamiltonian is diagonalized, 
it is easy to calculate the susceptibility using Eq. (\ref{eqn:chi'}). 
Since the unit cell contains three atoms, there are 
9 independent modes of polarization 
concerning the site degrees of freedom in the unit cell, 
and each of them has both spin-singlet and spin-triplet branches.
Correspondingly, the 
susceptibility is a $9\times9$ matrix at each ${\bf q}$, and 
its largest eigenvalue is the dominant response. 
As we are considering the noninteracting case ($U/t=0$) at the moment, 
the response of spin-singlet polarization is 
always degenerate with that of spin-triplet polarization. 
Of course, they start to differ, upon switching on the 
electron correlations, which will be discussed in a 
later part. 
In Fig. \ref{fig:EVchiU0}, we show the largest eigenvalue of 
the static spin (or charge) susceptibility $\hat{\chi'}({\bf q},0)$ 
in the Brillouin zone, for the isotropic case $t'/t=1.0$.
For the original Kagom\'e lattice, 
the susceptibility should have 
the exact six-fold rotational symmetry in the BZ. 
However, since in our treatment 
the original Kagom\'{e} lattice is 
transformed into the topologically-equivalent 
decorated square lattice ({\it isotropic Kagom\'e} lattice), 
the susceptibility has a distorted symmetry. 
Then the largest values of the susceptibility are located 
at ${\bf q}\approx \pm(0.4\pi, -0.4\pi)$, 
$\pm(0.4\pi, 0.8\pi)$, and $\pm(0.8\pi, 0.4\pi)$ 
and this is due to nesting behavior. 
Although there is not the strong nesting 
because of the rather isotropic Fermi surface, 
when the Fermi surface touches itself in a BZ 
at several points for a given nesting vector, 
this enhances the susceptibility. Such conditions are illustrated
in Fig. \ref{fig:Nesting} (a), where the arrows denote the 
corresponding nesting vectors. It is seen that these 
vectors indeed give rise to large values of the susceptibility.
The results for the anisotropic cases are also shown 
in the left panels of Fig. \ref{fig:chiS}, 
and they can also be explained by the Ferm surface nesting. 

\begin{figure}[tb]
\includegraphics[width=5cm]{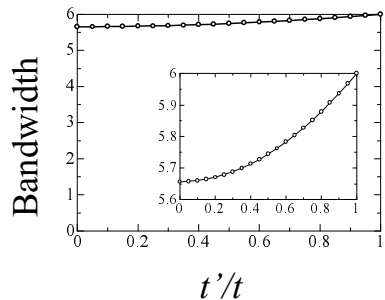}
\caption{The $t'$ dependence of the total bandwidth 
at non-interacting system. 
The inset shows the magnification.
}
\label{fig:anisotropy}
\end{figure}

In the following, we will study the effects of geometrical frustration 
of the isotropic/anisotropic Kagom\'e lattice 
with including the Coulomb interaction $U$. 
Of course, the strength of electron correlation is characterized 
by the ratio of the Coulomb interaction to the kinetic energy, 
but the latter is insensitive to the hopping anisotropy $t'/t$ 
and therefore the strength is essentially given by the ratio $U/t$. 
This is because the change in the total bandwidth 
and the DOS at the Fermi energy is small. 
We show the change in the bandwidth as a function of
$t'$ in Fig. \ref{fig:anisotropy}. 
Although the bandwidth decreases
slightly with decreasing hopping $t'$, its change is very 
small. Even at $t'/t=0$, the total 
bandwidth is $4\sqrt{2}t\approx 5.66t$, which is about 94\% of 
that of the isotropic case ($t'/t=1.0$). 
This enables us to systematically discuss electron correlations 
due to $U$ by using the anisotropic model.

\subsection{Magnetic properties}

\begin{figure*}[tb]
\includegraphics[width=16cm]{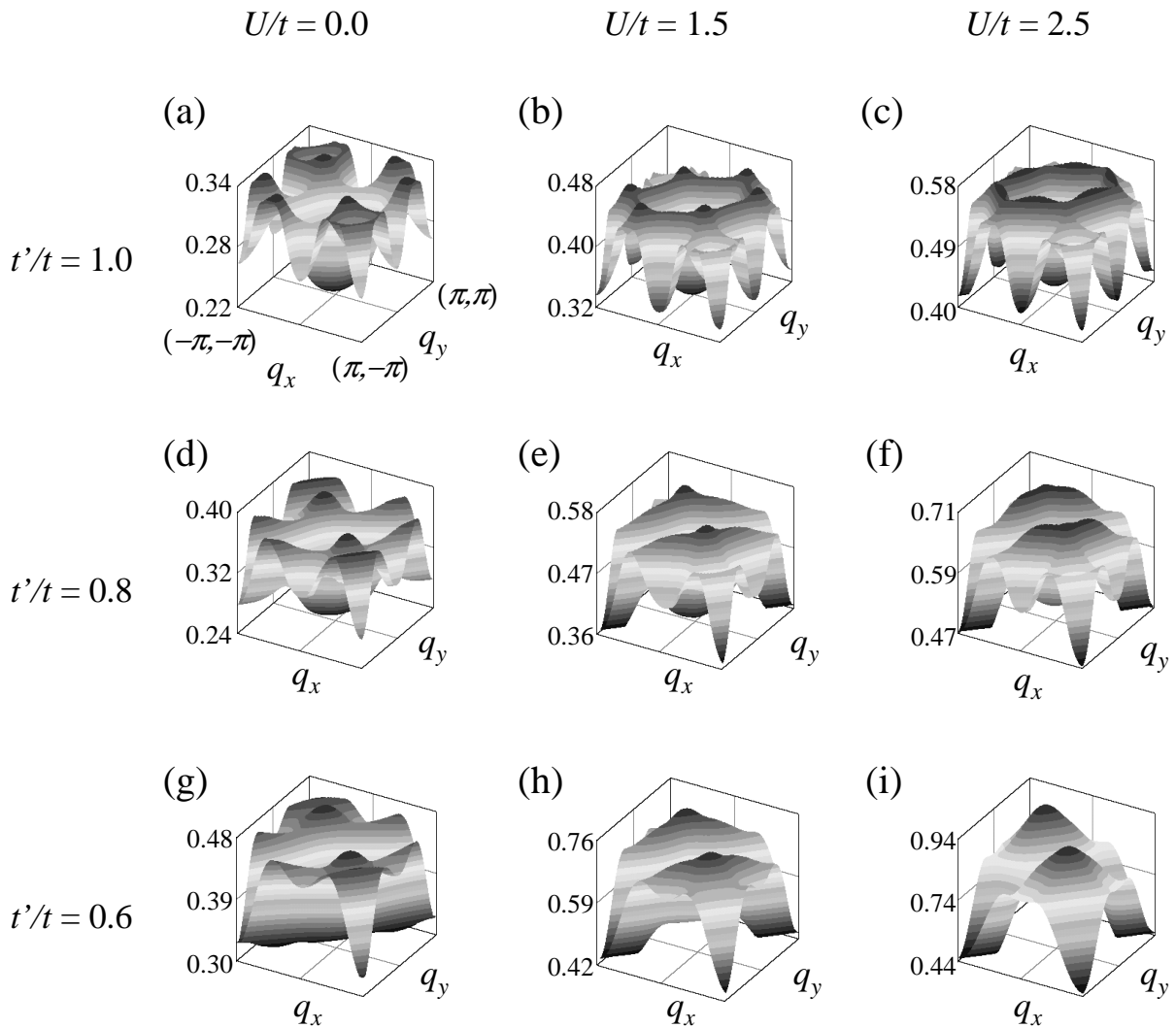}
\caption{The maximum eigenvalues of the static spin 
susceptibility $\hat{\chi}^{(s)}({\bf q},0)$ for three typical $U$ values. 
Upper, middle, and lower panels correspond to 
$t'/t=1.0$, $0.8$, and $0.6$, respectively. 
The temperature $T/t$ is 0.1. 
}
\label{fig:chiS}
\end{figure*}

We now turn to study electron correlations 
in the isotropic/anisotropic Kagom\'e lattice, 
and let us start with magnetic properties. 
Shown in Fig. \ref{fig:chiS} is the Coulomb interaction
 dependence of the maximum eigenvalues of the static spin 
susceptibility $\hat{\chi}^{(s)}({\bf q},0)$ for 
three typical values of $t'$. 
We first discuss the isotropic case ($t'/t=1.0$). 
For $U/t=0$, the largest values of the susceptibility
are located at the 6 points in the BZ, ${\bf q}\approx \pm(0.4\pi, -0.4\pi)$, 
$\pm (0.4\pi, 0.8\pi)$, and $\pm (0.8\pi, 0.4\pi)$. 
With increasing $U$, the susceptibility is enhanced in the whole BZ. 
However, its {\bf q}-dependence is reduced 
in contrast to the unfrustrated cases. 
In particular, it is found that the susceptibility for various {\bf q}'s 
besides the above-mentioned $6$ points are also strongly enhanced 
with increasing $U$, 
so that the {\bf q}-dependence of the spin susceptibility 
is weakened and spin fluctuations 
become more isotropic in the BZ. 
We also show the susceptibility for anisotropic cases 
$t'/t=0.8$ and $0.6$ in Fig. \ref{fig:chiS}. 
Although the (distorted) 
six-fold symmetry is lost due to hopping anisotropy, 
the spin susceptibility at $U/t=0$ still has large values 
at specific {\bf q}-positions, 
which are also able to be expected by the nesting picture 
similar to $t'/t=1.0$ case. 
When we switch on $U$, 
the peaks in the {\bf q}-dependence at $U/t=0$ gradually grows 
and this enhancement is more remarkable for smaller $t'$. 
For example, for $t'/t=0.6$, the peak positions 
in the intermediate $U$ region are 
${\bf q} \sim \pm (0.5\pi, -0.5\pi)$, 
which implies that the system prefers the spin configuration 
with the period of twice the original unit cell. 

We note here that just at $t'/t=0$, a flat band is located at the
Fermi level at half-filling. It is known that
 this gives rise to the so-called 
flat-band ferromagnetic ground state \cite{Lie89}. 
The nature of the ferromagnetism in this case has been well studied, 
so that we do not give detailed discussions for the cases $t'/t \ll 1.0$. 

\begin{figure}[tb]
\includegraphics[width=7cm]{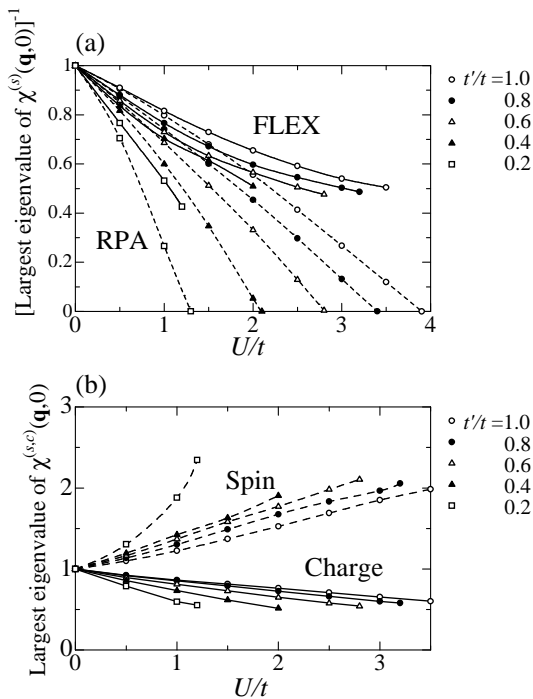}
\caption{(a) The inverse of the largest eigenvalue of the spin 
susceptibility as a function of the Coulomb interaction $U$ for 
various $t'$. Solid (dashed) lines represent the results of 
the FLEX approximation (RPA). 
(b) Largest eigenvalues of the spin and charge susceptibilities 
calculated by the FLEX approximation as a function of $U/t$. 
Solid (dashed) lines denote the charge (spin) sector, respectively. 
All of the data are normalized by the value at $U/t=0.0$. 
}
\label{fig:SDW&LVSC}
\end{figure}

In order to discuss the instability to spin ordered states, it is
instructive to compare the FLEX results with those calculated by 
the random phase approximation (RPA). 
The inverse of the largest eigenvalue of the spin 
susceptibility $\hat{\chi}^{(s)}({\bf q},0)$ in the whole BZ is shown 
in Fig. \ref{fig:SDW&LVSC} (a) for these two approximations. 
When this value becomes zero, 
the corresponding spin order, i.e., spin density wave (SDW), appears. 
Both results of the FLEX approximation and the RPA show 
that the dominant spin susceptibility is suppressed with increasing $t'$, 
indicating the suppression of spin order. 
Note that the RPA spin susceptibility diverges 
at a finite $U/t$ for all $t'$, and the divergence is even enhanced, 
as seen in the concave shape of the lines, due to the multi-band structure. 
On the other hand, the enhancement in the FLEX results 
is much weaker at large $U/t$ and the spin susceptibility 
does not seem to diverge. The behavior of the FLEX results is more 
reliable in the intermediate $U/t$ region, since mode-mode couplings 
are completely neglected in the RPA and the instability is overestimated. 
In particular, the suppression is most remarkable around the isotropic 
point $t' \sim t$, and we may conclude that 
no spin order is
realized for the isotropic Kagom\'e lattice within the FLEX approximation. 

Summarizing the magnetic properties, the spin susceptibility is enhanced 
in the presence of the Coulomb interaction in the whole BZ, 
but geometrical frustration of the Kagom\'e lattice 
has a strong effect to reduce the 
{\bf q}-dependence of the susceptibility and to suppress the instability to
any type of the SDW order. Although the present 
calculations are based on a weak coupling approach, the obtained 
results are quite consistent with properties expected in the strong 
coupling regime, the frustrated Heisenberg spin 
system, where the magnetic order is suppressed due to strong frustration. 

We have so far discussed the instability of spin-triplet polarization, 
but the instability in the spin-singlet channel is even weaker. 
The largest eigenvalues of the charge 
susceptibility $\hat{\chi}^{(c)}({\bf q},0)$ in the BZ are always smaller 
than that for $\hat{\chi}^{(s)}({\bf q},0)$ 
within the FLEX approximation and there is no indication of charge order 
(charge density wave, CDW)
or current-carrying state (shown in Fig. \ref{fig:SDW&LVSC} (b)). 
Thus we do not give detailed discussions on the charge susceptibility
here.

\subsection{One-Particle Spectral Function}
We now discuss the one-particle 
spectral function and the total spectral density, which are defined by
\begin{eqnarray}
A({\bf k},\omega)&=&-\frac{1}{3\pi}\sum_{m}
{\rm Im}G_{mm}({\bf k},\omega+i\delta)\\
\rho(\omega)&=&\frac{1}{N'}\sum_{\bf k}A({\bf k},\omega), 
\label{rho}
\end{eqnarray}
where $\delta$ is a small adiabatic constant. 

In order to obtain dynamical quantities in the FLEX approximation, 
the analytic continuation from the imaginary Matsubara frequency 
to the real frequency 
is usually performed by using the Pad\'e approximation 
or the maximum entropy method. However, it is 
known that this procedure sometimes encounters numerical difficulties. 
In the present study, we directly calculate the spectral function 
in a real frequency formulation without resorting to 
analytic continuation. 
(Our formulation is similar to those used 
in the previous works\cite{Dah95,Lan95,Wer96}, 
but simpler from the viewpoint of numerical techniques.) 
We need a small but finite adiabatic constant $\delta$ 
for the Green function, but its effects 
can be corrected systematically and 
we have checked that the Pad\'e approximation\cite{Vid77} 
reproduces similar results. 
We show here the results calculated by our real frequency technique, 
since they are free from numerical errors due to analytic continuation. 

\begin{figure}[tb]
\includegraphics[width=8cm]{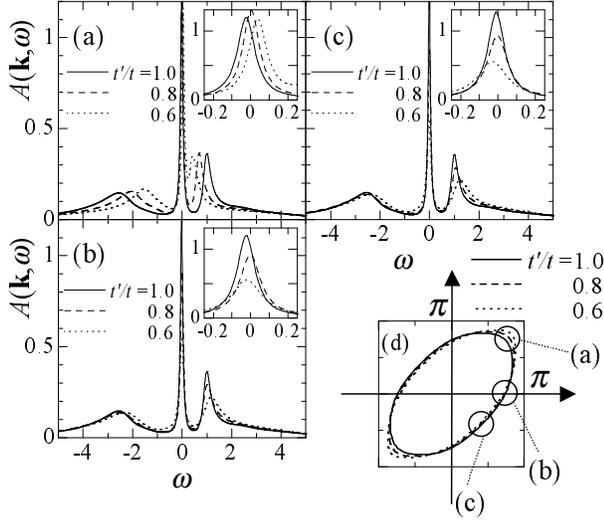}
\caption{One-particle spectra for various {\bf k}. 
(a) ${\bf k}$ $\approx$ ($0.75\pi,0.75\pi$), 
(b) ${\bf k}$ $\approx$ ($0.75\pi,0$), and 
(c) ${\bf k}$ $\approx$ ($0.42\pi,-0.42\pi$), respectively. 
(d) Fermi surface for several values of $t'$. 
Small circles represent the location of {\bf k} chosen for (a)-(c). 
The energy is measured from the Fermi level ($\omega=0$). 
The Coulomb repulsion $U/t=2.5$, the temperature $T/t=0.1$ 
and adiabatic constant $\delta=0.05$. 
The insets show the magnification of the low-frequency regime. 
}
\label{fig:Ak}
\end{figure}

Figure \ref{fig:Ak} shows the one-particle spectral function for 
several {\bf k}'s near the Fermi surface. 
All the spectra shown consist of three peaks 
reflecting the three-band structure; namely the
 $\alpha$-th peak is mainly due to the contribution of the 
 band with $E_{{\bf k}\alpha}$. As $U$ increases, 
two peaks at large $|\omega|$ is gradually smeared due
to life-time effects.
On the other hand, the peak near $\omega=0$ keeps a sharp 
structure, implying a well-defined Fermi liquid behavior. 
This behavior is particularly prominent in the 
isotropic case $t'/t=1.0$ irrespective of {\bf k}'s, (a)-(c). 
However in the anisotropic cases of 
$t'/t=0.8$ and $0.6$, 
the central peak changes its shape 
from (a) ${\bf k}\approx (0.75\pi,0.75\pi)$ to 
(c) ${\bf k}\approx (0.42\pi,-0.42\pi)$ 
where the amplitude of the spectra at $\omega \sim 0$ is reduced 
with the decrease of $t'/t$. 
This behavior suggests that quasi-particles
are stabilized in the region of isotropic hopping $t'\sim t$. 
As discussed in the previous subsection, 
increasing $t'$ represents the enhancement of 
effects of geometrical frustration, 
and this leads to the trend that the wavevector dependences 
are strongly reduced in the BZ. 
Since the spin fluctuations are localized in the real space 
and these amplitude are strongly reduced, 
various orders, such as 
the SDW, CDW, and currant-carrying states, are suppressed 
and the metallic state is stabilized. 
Therefore the quasi-particle picture becomes fine description. 
This result indicates that geometrical frustration on the Kagom\'e lattice 
may be important for the formation of quasi-particles. 

\begin{figure}[tb]
\includegraphics[width=6cm]{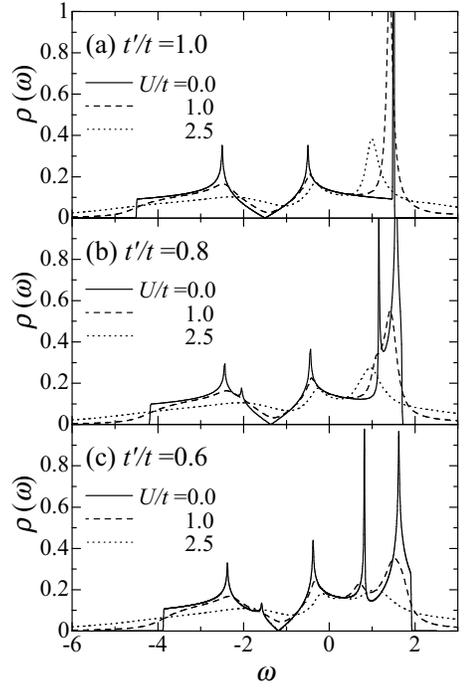}
\caption{Total spectral density for 
various $U$. (a) $t'/t=1.0$, (b) $t'/t=0.8$ and
 (c) $t'/t=0.6$. 
}
\label{fig:LD}
\end{figure}

For reference, we also present in Fig. \ref{fig:LD} the results of 
the total spectral density defined by Eq. (\ref{rho}). 
As $U$ increases, the total spectral density is smeared in the whole energy 
region due to the 
life-time effect. However, in the weak-coupling regime treated here, 
the total spectral density near the Fermi level is rather 
insensitive to the change in 
anisotropic hopping $t'$ in comparison with other physical quantities, 
such as the one-particle spectra, etc. 
Although the one-particle spectral weight with {\bf k} 
for the $k_{y}=-k_{x}$ direction close to the Fermi surface 
is strongly reduced at $\omega \sim 0$ with decreasing $t'/t$, 
the amplitude of total spectral density at $\omega \sim 0$ 
becomes largest at $t'/t=0.6$ even when $U$ is introduced. 
In the intermediate $U$ region, 
the one-particle spectrum at {\bf k} 
far from the Fermi surface mainly consists of 
the incoherent parts. 
Since the local spectral density is obtained by 
the {\bf k}-summation in Eq. (\ref{rho}), 
the contributions of these incoherent parts far from Fermi surface 
become quite large even at $\rho(0)$ with increasing $U$. 
Therefore, it is difficult to discuss about 
the coherence of quasi-particles 
only based on the total spectral density $\rho(0)$, 
and we need more detailed information from $A({\bf k}, \omega)$.

Summarizing this subsection, as compared with the total spectral density, 
the one-particle spectrum gives us more detailed information 
on the nature of quasi-particles. 
The weight of the coherent part in the one-particle spectrum 
grows with increasing $t'$, 
which means that the quasi-particle behavior is stabilized 
by geometrical frustration.

\subsection{Renormalization effects}

To investigate the properties of quasi-particles in more detail,
we calculate the renormalization factor and the Fermi 
velocity of quasi-particles. 
According to the conventional Fermi liquid theory, the one-particle 
spectrum consists of the $\delta$-function like coherent part 
and the broad incoherent part. As the Coulomb interaction increases, 
the quasi-particle 
mass is enhanced and the weight of the coherent part becomes small.
For the single band case, the renormalization factor 
is given by the self energy as 
\begin{eqnarray}
Z_{\bf k}=\bigg[1-\frac{\partial {\rm Re} \Sigma ({\bf k},\omega)}{\partial \omega}\bigg|_{\omega=0}\bigg]^{-1}, 
\end{eqnarray}
which corresponds to the weight of the coherent part. 
However, in the present case, the 
one-particle spectrum at a given {\bf k} has contribution of three bands. 
When the coherent quasi-particle part is separated from 
the other parts, it is given by 
\begin{eqnarray}
A^{\rm coh}({\bf k},\omega)=\frac{1}{\pi}\frac{Z_{\bf k}\gamma}
{(\omega-\xi_{\bf k})^{2}+\gamma^{2}}+b, 
\label{eqn:Akcoh}
\end{eqnarray}
where $\xi_{\bf k}$ is the shift from the Fermi level, 
$b$ is the correction due to 
the other peaks which are located far from the Fermi level, 
and $\gamma$ represents the inverse of life-time of the quasi-particle. 
These parameters can be determined by numerical fitting 
of the data of the spectral function. 
However instead of performing this procedure, 
we here use an alternative approach to simply determine 
the renormalization factor. 
Since the middle band near the Fermi level is 
isolated sufficiently far from the other bands, 
we can introduce the following 
effective self-energy in the low-energy region. 
\begin{eqnarray}
\tilde{\Sigma}^{\rm eff}({\bf k},\omega)&\equiv& \tilde{g}
({\bf k},\omega)^{-1}-\tilde{G}({\bf k},\omega)^{-1}, 
\label{eqn:ApproxSigma}
\end{eqnarray}
where 
\begin{eqnarray}
\tilde{g}({\bf k},\omega)&=&\frac{1}{3}\sum_{m}{g}_{mm}({\bf k},\omega),\\
\tilde{G}({\bf k},\omega)&=&\frac{1}{3}\sum_{m}{G}_{mm}({\bf k},\omega). 
\end{eqnarray}
The approximate formula for the renormalization factor then reads,
\begin{eqnarray}
Z_{\bf k}&=&\bigg[1-\frac{\partial {\rm Re}\tilde{\Sigma}^{\rm eff}({\bf k},\omega)}{\partial \omega}\bigg|_{\omega=0}\bigg]^{-1}.
\label{eqn:ApproxZk}
\end{eqnarray}
Although Eq. (\ref{eqn:ApproxZk}) is not exact, this 
formula works well as far as the low energy regime is concerned. 
We have checked for several {\bf k} that
the results obtained by using Eq. (\ref{eqn:ApproxZk}) 
are in good agreement with the renormalization factor 
determined by the fitting Eq. (\ref{eqn:Akcoh}). 

\begin{figure}[b]
\includegraphics[width=8cm]{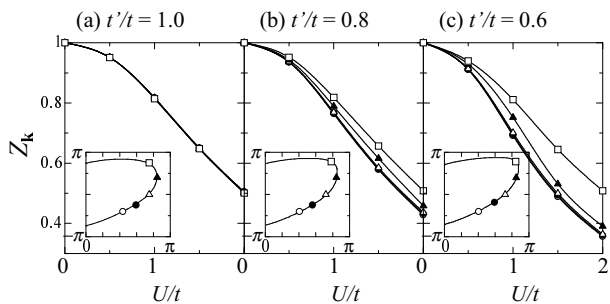}
\caption{Renormalization factor as a function of $U$ at 
temperature $T/t=0.1$; (a) $t'/t=1.0$, (b) $t'/t=0.8$, and (c) $t'/t=0.6$, 
respectively. The insets show the Fermi surface (solid line) in 
the half BZ ($0\leq k_{x}\leq\pi$) where symbols (open circles etc.) 
denote various {\bf k} for which $Z_{\bf k}$ is calculated. 
}
\label{fig:Zk}
\end{figure}

Figure \ref{fig:Zk} shows the {\bf k}-dependence of the renormalization 
factor as a function of $U$ for various $t'$. 
We show the results only in the 
{\bf k} region, $0\leq|k_{y}|\leq k_{x}\leq \pi$, because the values in 
other regions are readily obtained according to the symmetry property. 
In all the cases of $t'/t$, 
the renormalization factor $Z_{\bf k}$ 
decreases with the increase of $U$, 
meaning that the effective mass is enhanced. 
An important point is that 
$Z_{\bf k}$ becomes nearly independent 
of {\bf k} in the isotropic case ($t'/t=1.0$), being consistent with 
the results of one-particle spectral functions. 
On the other hand, with decreasing $t'$, the 
difference in $Z_{\bf k}$ for different {\bf k}'s 
becomes more remarkable. 
In particular, the $Z_{\bf k}$ for 
{\bf k}'s along the direction of $k_{y}=-k_{x}$ is strongly renormalized. 
It should be noted that 
the largest $Z_{\bf k}$ in the anisotropic case 
(open squares in Fig. \ref{fig:Zk} (b) and (c)) 
has almost the same amplitude of that for $t'/t=1.0$. 
In other words, with decreasing $t'$, 
the quasi-particle weight is reduced in comparison with that of $t'/t=1.0$.
Therefore this is another evidence that 
geometrical frustration stabilizes the metallic state 
up to larger $U$ regime 
when $t'/t=1.0$ (the isotropic Kagom\'e lattice). 

\begin{figure}[b]
\includegraphics[width=8cm]{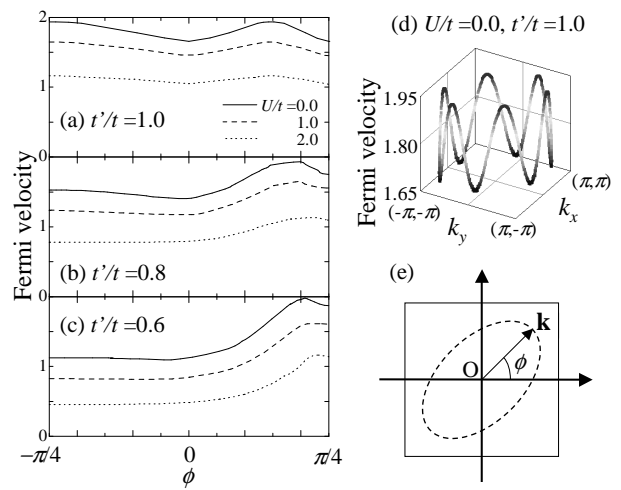}
\caption{Angular dependence of the Fermi velocity for various $U$; 
(a) $t'/t=1.0$, (b) $t'/t=0.8$, and (c) $t'/t=0.6$ (Eq. (\ref{lVkl})). 
(d) Fermi velocity at $U/t=0$ and $t'/t=1.0$ in the {\bf k} space. 
(e) angle $\phi$ of the {\bf k} point on the Fermi surface. 
}
\label{fig:Vk}
\end{figure}

Shown in Fig. \ref{fig:Vk} is the absolute value of the Fermi 
velocity, which is given by 
\begin{eqnarray}
|v^{F}_{\bf k}|&=&\sqrt{\bigg(\frac{\partial \tilde{E}_{\bf k}}
{\partial k_{x}}\bigg)^{2}+\bigg(\frac{\partial \tilde{E}_{\bf k}}
{\partial k_{y}}\bigg)^{2}}, 
\label{lVkl}
\end{eqnarray}
which is obtained from the renormalized quasi-particle 
energy of the middle band, 
\begin {eqnarray}
\tilde{E}_{\bf k}&=&Z_{\bf k}\bigg(E^{(2)}_{\bf k}-\mu+{\rm Re}
\tilde{\Sigma}^{\rm eff}({\bf k},0)\bigg). 
\end{eqnarray}
In the case of $t'/t=1.0$, the Fermi velocity has 
the distorted six-fold rotational symmetry in the BZ reflecting the 
lattice structure for (Fig. \ref{fig:Vk} (d)) 
and the ratio of the largest and smallest velocity 
is about $1.17$ for $U/t=0$. 
With increasing $U$, the ratio 
is gradually reduced to the smaller value of about $1.10$ for $U/t=2.0$, 
indicating again the suppression of the {\bf k}-dependence. 
On the other hand, for the anisotropic case $t'/t<1.0$, 
the ratio is not reduced but enhanced. 
For instance, for $t'/t=0.6$, 
the ratio changes from about $1.79$ for $U/t=0.0$ 
to $2.45$ for $U/t=2.0$. 
These results are in accordance with 
those of the renormalization factor in the sense that 
the wavevector dependence grows with increasing hopping anisotropy. 

\section{Summary and Discussions}
In this paper, we have studied how geometrical frustration 
of the isotropic/anisotropic Kagom\'e 
lattice affects physical properties in a metallic phase. 
By using the FLEX approximation for the Hubbard model 
at half-filling, we have calculated the spin and charge susceptibility, 
the one-particle spectrum, the total spectral density, 
the quasi-particle renormalization factor, and the Fermi velocity. 
In the isotropic Kagom\'e lattice, we have shown that 
the spin susceptibility is 
enhanced by the Coulomb interaction as far as the amplitude 
of susceptibility is concerned. However, in contrast to ordinary 
cases of magnetic instability, its wavevector dependence is 
strongly suppressed at the same time, and there appear many peaks in the BZ
accumulating to form a line. 
This reduction of the wavevector dependence around the Fermi surface 
is also observed for other quantities, such as 
the quasi-particle weight of the one-particle spectrum. 
We have demonstrated that this behavior is indeed due to the 
frustrated lattice geometry by controlling frustration 
by anisotropic hopping; 
the {\bf k}-dependence is considerably recovered 
with the decrease of $t'$ i.e. when frustration is reduced. 
By comparing the spin susceptibility calculated by the 
FLEX approximation with that of RPA, we have shown
that the instability to magnetically ordered phases 
is dramatically suppressed by mode-mode couplings of 
fluctuations, which are neglected in the RPA level. 
This suppression is most prominent in the isotropic 
case, i.e. most frustrated case. 
Furthermore, considering the one-particle spectrum, 
we have shown that the coherent part has a large weight 
in the isotropic Kagom\'e lattice up to larger $U$ regime, 
which is also consistent with the behaviors of 
the renormalization factor and Fermi velocity. 
These results indicate that 
geometrical frustration stabilizes the formation of quasi-particles. 

The main conclusion of the present study is that 
the strong geometrical frustration reduces 
the wavevector dependence of spin fluctuations 
and suppresses the instabilities to various orders, 
such as the SDW, CDW, and current-carrying states, 
so that the metallic state is stabilized 
and the quasi-particles become well-defined 
in the isotropic Kagom\'e lattice. 
This conclusion derived from the FLEX approximation, 
which is a essentially weak-coupling approach, 
but it is consistent with the results known for the strong coupling 
model, i.e., the absence of any magnetic order due to strong frustration 
in the antiferromagnetic spin system on the Kagom\'e lattice. 
Therefore, it is reasonable that characteristic 
properties discussed in the present work for the weak-coupling
Hubbard model on the 
Kagom\'e lattice are naturally connected to those inherent in 
the frustrated Heisenberg spin system. 

Within the FLEX approximation in this paper, 
we have not observed other ordered states, 
either such as the staggered flux (current-carrying) state 
\cite{Kot88,Aff88,Nay02} and 
the charge ordered state\cite{Wer41} etc., 
which are other possible orders in the strong coupling regime. 
In the present case of half-filling, 
although charge ordered states are unlikely, 
a staggered flux state may appear 
due to strong geometrical frustration. 
Furthermore, upon finite hole doping, 
geometrical frustration may realize a charge ordered state. 
It is an interesting open problem to be addressed 
in the future.

Experimentally, most compounds studied so far
have been restricted to spin systems on 
the Kagom\'e lattice, such as ${\rm SrCr_{8}Ga_{4}O_{19}}$ 
\cite{Ram90,Uem94}. 
It is thus interesting 
to study electron correlations in metallic Kagom\'e lattice 
materials.

\section*{Acknowledgements}
One of the authors (YI) would like to acknowledge helpful discussion 
with T. Takimoto. The work is partly supported by a Grant-in-Aid 
from the Ministry of 
Education, Science, Sports, and Culture of Japan. 
A part of the numerical calculations were performed on computers 
at the supercomputer center 
at the Institute for the Solid State Physics, The University of Tokyo, 
and at Yukawa Institute Computer Facility, Kyoto University.


\end{document}